\documentclass[a4paper,aps,prd,twocolumn,nofootinbib]{revtex4}
\usepackage[latin1]{inputenc}
\usepackage{amsfonts}
\usepackage{subfigure}
\usepackage{epsfig}
\usepackage{geometry}
\usepackage[dvips]{color}
\usepackage{amssymb}
\usepackage{amscd}
\usepackage[mathscr]{eucal}
\usepackage{latexsym}
\usepackage{amsmath}
\DeclareMathAlphabet{\scr}{U}{rsfs}{m}{n}
\newcommand{\newc}{\newcommand}
\newc{\beq}{\begin{equation}}
\newc{\eeq}{\end{equation}}
\newc{\beqn}{\begin{eqnarray}}
\newc{\eeqn}{\end{eqnarray}}
\DeclareMathAlphabet{\scr}{U}{rsfs}{m}{n}
\begin{document}
\pagestyle{plain}
%
%\date{\today}
~

\hfill {\footnotesize Preprint number CERN-PH-TH/2007-042}
\title{{\bf{~\\\Large Supersymmetric Jarlskog Invariants: the Neutrino Sector\\~}}}
\author{Herbi K. Dreiner\,\raisebox{0.1cm}{\tiny 1}}\email[E-mail:\ ]
{dreiner@th.physik.uni-bonn.de}
\author{Jong Soo Kim\,\raisebox{0.1cm}{\tiny 1}}\email[E-mail:\ ]
{jsk@th.physik.uni-bonn.de}
\author{Oleg Lebedev\,\raisebox{0.1cm}{\tiny 2}}\email[E-mail:\ ]
{oleg.lebedev@cern.ch}
\author{Marc Thormeier\,\raisebox{0.1cm}{\tiny 3}}\email[E-mail:\ ]
{marc.thormeier@cea.fr}
\affiliation{
~\\$^1$Physikalisches Institut der Universit\"at Bonn,\\ 
  Nu\ss{}allee 12, 53115 Bonn, Germany\\
~\\$^2$CERN, PH-TH,  CH-1211 Geneva 23, Switzerland\\
~\\$^3$Service de Physique Th\'eorique, CEA-Saclay,\\
  Orme des Merisiers, 91191 Gif-sur-Yvette Cedex, France\\}
\begin{abstract} \noindent
We generalize the notion of the Jarlskog invariant to supersymmetric
models with right--handed neutrinos.  This allows us to formulate
basis--independent necessary and sufficient conditions for CP
conservation in such models.
\end{abstract}
\maketitle
%
%%%%%%%%%%%%%%%%%%%%%%%%%%%%%%%%%%%%%%%
%

\section{Introduction}

CP violation in the quark sector of the Standard Model (SM) is
controlled by the Jarlskog invariant \cite{Jarlskog:1985ht}, 
\begin{equation}
{\rm Im} \Bigl(  {\rm Det}[Y^u Y^{u\dagger}, Y^d Y^{d\dagger} ] 
\Bigr) \;,
\end{equation}
which can also be written in the form \cite{Bernabeu:1986fc},\cite{Gronau:1986xb}
\begin{equation}
{\rm Im} \Bigl(  {\rm Tr} [ Y^u Y^{u\dagger}, Y^d Y^{d\dagger} ]^3  \Bigr) \;,
\end{equation}
where $Y^{u,d}$ are the quark Yukawa matrices.  This is a CP--odd
quantity, invariant under quark basis transformations.  CP violation
is possible if and only if the Jarlskog invariant is non--zero
(assuming $\bar \theta_{\rm QCD}=0$). This is a simple and powerful
result.

In the lepton sector, the situation is more complicated. Assuming that
the smallness of the neutrino masses is explained by the seesaw
mechanism \cite{Minkowski:1977sc}-\cite{Mohapatra:1979ia}, the
effective neutrino mass matrix is of the Majorana type.  It has
different basis transformation properties compared to the Dirac case.
This results in three independent CP phases and more complicated
CP--odd invariants \cite{Branco:1986gr}. A recent discussion of this
subject is given in \cite{Branco:2004hu}. Applications of the
invariant technique to physics beyond the SM can be found in
\cite{Botella:1985gb}-\cite{Gunion:2005ja}.

A generalization of the Jarlskog invariant to supersymmetric models
was constructed in \cite{Lebedev:2002wq}. It was found that CP
violation is controlled in this case by a different type of invariants
containing an antisymmetric product of three flavour matrices.
Applications of this approach were studied in \cite{Botella:2004ks}.
In this work, we extend these results to SUSY models with
right--handed neutrinos. As seen in the SM case, this brings in
flavour objects with ``unusual'' transformation properties and leads
to distinct physics.

In what follows, we first study CP--phases and invariants in the SM
with three right--handed neutrinos. We differ from previous work
in implementing the concise techniques of \cite{Lebedev:2002wq}. Within
 this formalism, we then construct the SUSY generalization, the
Minimal Supersymmetric Standard Model (MSSM) with three right--chiral
neutrino superfields, and give an example of possible applications.

\section{ SM With Three Right-Handed Neutrinos\label{SM}}\noindent 

Consider an extension of the SM with three right-handed neutrinos. 
The  relevant terms in the leptonic  Lagrangian density are
\begin{eqnarray}\label{SMlagrangian}
\Delta \scr{L}&=&  
Y^{e}_{ij}\bar l_i e_j \mathcal{H} + Y^{\nu}_{ij} \bar l_i \nu_j 
\widetilde{\mathcal{H}}\nonumber +
\frac{1}{2}M_{ij}\bar\nu_i^c\nu_j+\mathrm{H.c.},
\end{eqnarray}
where $l$, $e$, $\nu$ and $\mathcal{H}$ denote the left-handed charged
lepton doublet, the right-handed charged lepton singlet, the
right-handed neutrino singlet and the Higgs doublet, respectively.
$\widetilde{\mathcal{H}}$ is given by $i\tau_2 \mathcal{H}^*$, where
$\tau_2$ is the second Pauli matrix.  $Y^e_{ij}$ is the charged lepton
Yukawa matrix, $Y^\nu_{ij}$ is the Yukawa matrix for the neutrinos,
and $M_{ij}$ is the complex symmetric Majorana mass matrix for the
right-handed neutrinos. $i,\, j$ are the generation indices and the
superscript~$c$ denotes charge conjugation.

The kinetic terms are invariant under unitary basis transformations
\begin{eqnarray}\label{symmetries}
U(3)_l \times U(3)_e \times U(3)_\nu \;,
\label{tot-sym}
\end{eqnarray} 
namely 
\begin{eqnarray}
l&\rightarrow&U_l^\dagger  ~ l\label{transformation_doublet},\\
e&\rightarrow&U_e^\dagger ~ e\label{transformation_singlet},\\
\nu&\rightarrow&U_\nu^\dagger ~ \nu\label{transformation_neutrino}.
\end{eqnarray}
This means that a theory with the flavour matrices
transformed according to
\begin{eqnarray}
Y^e &\rightarrow& U_l^\dagger ~Y^e~ U_e,\label{transformation_Ye}\\\
Y^\nu&\rightarrow&U_l^\dagger ~Y^\nu~ U_\nu,\label{transformation_Ynu}\\
M&\rightarrow&U_\nu^T ~M~ U_\nu\label{transformation_M}
\end{eqnarray} 
represents the same physical situation and is equivalent to the
original one. With an appropriate choice of the phase convention, the
CP operation amounts to complex conjugation of these matrices (see
e.g.\cite{delAguila:1995bk}),
\begin{equation}
{\cal M} \rightarrow {\cal M}^* \;,
\end{equation}
where ${\cal M}= \{ Y^e, Y^\nu, M    \} $. If this operation can be ``undone''
by a symmetry transformation,  no CP violation is possible. 

Physical CP violation is controlled by CP--violating basis independent
invariants {\it \`a la} Jarlskog. This allows one to formulate
necessary and sufficient conditions for CP conservation in a basis
independent way. On the other hand, it is also instructive to study CP
violating phases in a specific basis, taking advantage of symmetries
of the system.  In what follows, we will pursue both of these
approaches.

In seesaw models, the scale of the Majorana mass matrix is taken to be 
very large, around  the GUT  scale. In this case, the low energy theory
is obtained by integrating out the right--handed neutrinos. This produces
a dimension-5 operator involving the left--handed leptons  and
an effective coupling constant
\begin{eqnarray}
m_{\mathrm{\mathrm{eff}}}=Y^\nu M^{-1} Y^{\nu^T}  \;,
\label{meff}
\end{eqnarray}
which results in  neutrino masses  upon electroweak symmetry breaking.
The apparent flavour symmetry of this low energy theory is
\begin{equation}
U(3)_l \times U(3)_e
\end{equation}
with the transformation law
\begin{eqnarray}
Y^e&\rightarrow& U_l^\dagger Y^e U_e \;, \nonumber  \\
m_{\mathrm{eff}}&\rightarrow& U_l^\dagger m_{\mathrm{eff}} U_l^*\;.
\label{transformation_meff}
\end{eqnarray} 

The number of independent CP phases can be obtained by a
straightforward parameter counting.  In the high energy theory, $Y^e,
Y^\nu $ and $M$ contain $9+9+6=24$ phases. A unitary $3\times 3$
matrix representing basis transformations has 6 phases, which means
that 18 phases can be removed.\footnote{If the Majorana mass matrix
were absent, only 17 phases could be removed since a phase
transformation proportional to the unit matrix leaves $Y^e$ and
$Y^\nu$ intact, which corresponds to a conserved lepton number.  } 
Thus we end up with six physical phases at high energies. In the low
energy theory, $Y^e$ and $m_{\rm eff}$ contain $9+6=15$ phases. 12 of
them can be removed by unitary transformations, while three are
physical. Clearly, the other three physical phases of the high energy
theory are associated with the heavy neutrinos and cannot be observed
at low energies.  However, these can be relevant to CP violation at
high energies, \emph{e.g.} leptogenesis \cite{Davidson:2003yk}.

In what follows, we study in more detail these CP phases and the
corresponding invariants.

\subsection{High--Energy Theory}
\subsubsection{CP phases}

Let us first identify the physical CP phases in a specific basis
assuming a general form of $Y^e, Y^\nu $ and $M$.  The unitary
transformations (\ref{transformation_Ye}-\ref{transformation_M}) allow
us to bring the flavour matrices into the form
\begin{eqnarray}
&& Y^{\nu}= {\rm real~diagonal}\;,\; \nonumber \\ 
&& Y^{e}= {\rm Hermitian}\;,\; \label{specbasis}\\
&& M= {\rm symmetric } \;, \nonumber 
\end{eqnarray}
where the last equation is  satisfied in any basis.
This basis is defined only up to a diagonal phase transformation
\begin{equation}
\widetilde U_l=\widetilde U_e= \widetilde U_\nu  =\mathrm{diag}(\exp[i\alpha_1],
\exp[i\alpha_2],\exp[i\alpha_3])\,. 
\label{phaseredef}
\end{equation}
Under this residual symmetry, $Y^e$ and $M$ transform as 
\begin{eqnarray}
Y^e_{ij}&\rightarrow& Y^e_{ij}~\exp[\mathrm{i}(\alpha_j-\alpha_i)],
\label{phase_transformation_H}\\
M_{ij}&\rightarrow& M_{ij}~\exp[\mathrm{i}(\alpha_i+\alpha_j)].
\label{phase_transformation_M}
\end{eqnarray}
The physical CP phases must be invariant under these  transformations. 
Since $Y^{e}$ and $M$ have 9 phases, only 6 of them  are are physical.

The simplest invariant CP phase is a CKM--type phase which is the only
one surviving  the limit $M\rightarrow 0$. It is given by
\begin{equation}
\phi_0=\mathrm{arg}[Y^e_{12}Y^e_{23}Y^{e*}_{13}]\;.\label{phase0}
\end{equation}
The other five phases involve $M$. Three of them can be built
entirely out of $M$,
\begin{eqnarray}
\phi_1&=&\mathrm{arg}[M_{11}M_{22}M_{12}^{*2}]\;,\label{phase1}\\
\phi_2&=&\mathrm{arg}[M_{22}M_{33}M_{23}^{*2}]\;,\label{phase2}\\
\phi_3&=&\mathrm{arg}[M_{11}M_{33}M_{13}^{*2}] \;,\label{phase3}
\end{eqnarray}
while the remaining two involve $Y^e$ as well,
\begin{eqnarray}
\phi_4&=&\mathrm{arg}[Y^e_{13}M_{13}M_{33}^*],\label{phase4}\\
\phi_5&=&\mathrm{arg}[Y^e_{23}M_{22}M_{23}^*].\label{phase5}
\end{eqnarray}
It should be clear by considering the independent matrix entries, that
these phases are independent.

The necessary and sufficient conditions for CP conservation are given by
\begin{equation}
\phi_i=0 
\label{phi=0}
\end{equation}
for $i=0,..,5$, and the phases are understood \mbox{mod~$\pi$.} If
these conditions are satisfied, the flavour objects in
Eq.~(\ref{specbasis}) can be made real by choosing appropriate
$\alpha_i$. Then no CP violation is possible. Conversely, CP
conservation implies that the flavour matrices are real in some
basis. Then, the CP conserving $Y^e, Y^\nu $ and $M$ are generated by
the phase redefinitions in (\ref{phaseredef}), leaving $\phi_i=0$
intact.

\subsubsection{CP Violating Invariants}

Conditions for CP conservation can also be formulated in a basis independent way.
To do that, one first forms matrices which are manifestly invariant under two of
the unitary symmetries, then builds CP--odd traces  out of them.

Consider the following Hermitian matrices
\beqn 
A &\equiv& Y^{\nu\dagger} Y^\nu,\label{eq:aequiv}\\
B&\equiv& Y^{\nu\dagger} Y^e Y^{e\dagger} Y^\nu, \\
C&\equiv& M^*M,\\
D&\equiv& M^* (Y^{\nu\dagger} Y^\nu)^*M.  \label{eq:aequiv1}
\eeqn
In general, they are not diagonalizable simultaneously and transform as 
\begin{eqnarray}
 {\cal M}_i  &\rightarrow& U_\nu^\dagger ~ {\cal M}_i ~ U_\nu \;, 
\end{eqnarray}
where ${\cal M}_i= \{ A,B,C,D \}$. The simplest CP-odd invariants that
can be formed out of this set are
\begin{eqnarray}
&& {\rm Tr} [ {\cal M}_i^p , {\cal M}_j^q]^n \;, \nonumber\\
&& {\rm Tr} [  {\cal M}_i^p , {\cal M}_j^q , {\cal M}_k^r ]^m \;, 
\end{eqnarray}
where $p,q,r$ are integer and $n,m$ are odd; $ [...] $ denotes
complete antisymmetrization of the matrix product.  The first class
(``$J$--type'') of invariants is the familiar Jarlskog type, while the
second class (``$K$--type'') appears, for example, in supersymmetric
models \cite{Lebedev:2002wq}, see also Eqs.~(\ref{jk}) below.  These
objects are CP-odd since the CP operation on the fields is equivalent
to complex conjugation of the matrices, which is in turn equivalent to
a transposition for Hermitian matrices. In a specific basis [for
instance, (\ref{specbasis})], these objects are functions of the six
physical CP phases. In the non--degenerate case which we are
considering, the vanishing of six independent invariants implies the
vanishing of the physical CP phases. This means in turn that all
possible CP violating invariants are zero and CP is conserved.

An admissible choice of independent invariants is\footnote{We drop the
Im(...)  for each invariant in the following.}
\begin{eqnarray}
&&\mathrm{Tr}[A,B]^3,\label{j0}\\
&&\mathrm{Tr}[A,C]^3,\label{j1}\\
&&\mathrm{Tr}[A,D]^3,\label{j2}\\
&&\mathrm{Tr} ([A,C]B ),\label{j3}\\
&&\mathrm{Tr}([A,D]B),\label{j4}\\
&&\mathrm{Tr}([A,D]C),\label{j5}
\end{eqnarray}
where we have used Tr$[a,b,c] \propto $ Tr$[a,b]c$. The first
invariant is proportional to the sine of the CKM--type phase $\phi_0$,
while the others depend in a complicated way on all of the phases
(\ref{phase0})-(\ref{phase5}). It is a non--trivial task to determine
whether given invariants are mutually independent. To do that, we
calculate the Jacobian
\begin{equation}
{\rm Det} \Bigl(   \frac{\partial \scr{J}_i}{\partial \phi_j }   \Bigr) \;,
\label{J}
\end{equation} 
where $\scr{J}_i$ are the invariants above. A non--zero Jacobian
indicates that the objects are independent. We confirm that this is
indeed the case.

It is instructive to consider the above invariants in a specific basis,
for example, where matrix $A$ is diagonal,
\begin{equation}
A={\rm diagonal} \;.
\end{equation}
This basis is defined up to a rephasing
\begin{equation}
\tilde U_\nu= \mathrm{diag}(\exp[i\alpha_1],\exp[i\alpha_2],
\exp[i\alpha_3])\;.
\end{equation}
The physical CP phases must  be  invariant under this residual
symmetry and are of the form  
\begin{eqnarray}
&& {\rm arg} [B_{12} B_{23} B_{13}^* ]~,~  {\rm arg} [C_{12} C_{23} C_{13}^* ]~,... 
\label{arg1}\\
&& {\rm arg} [B_{12} C_{12}^*]~,~{\rm arg} [B_{23} C_{23}^*]~,...\label{arg2}
\end{eqnarray}
For $N$ independent Hermitian objects one can form $3N-5$ independent
invariant phases and all of the invariants depend on these $3N-5$
variables.  This can be understood by parameter counting: $N$
Hermitian matrices contain $3N$ phases and unitary basis
transformations $U_\nu$ absorb $6-1=5$ of them since the overall phase
transformation leaves all the matrices intact. The explicit dependence
of the invariants on these phases has been studied in
\cite{Lebedev:2002wq}.

In our case, there appear to be seven phases according to this
argument. However, not all of our Hermitian matrices are completely
independent as they are built out of three flavour matrices. One of
the phases is a function of the others and we have six truly
independent CP phases as explained in the previous subsection. These
are rather complicated functions of the expressions
(\ref{arg1}) and (\ref{arg2}), except
\begin{equation}
\phi_0 \propto {\rm arg} [B_{12} B_{23} B_{13}^* ] \;.
\end{equation}
Note that if we chose only three Hermitian matrices $A,B,C$ to work
with, we could only extract four CP phases regardless of how many
invariants we would write. So, some information is lost when
constructing Hermitian objects.  It is thus necessary to include a
further matrix $D$, which brings in additional input. To show that
this is sufficient, one must calculate the Jacobian (\ref{J}).

The necessary and sufficient conditions for CP conservation in the
non--degenerate case are
\begin{equation}
\scr{J}_i=0 \;,
\end{equation}
where $\scr{J}_i$ are the invariants (\ref{j0})-(\ref{j5}). This is
equivalent to Eq.(\ref{phi=0}).

\subsection{Low--Energy Theory
\label{lowhighSM}}
\subsubsection{CP Phases}

At low energies, we have two flavour matrices $Y_e$ and $m_{\mathrm
{eff}}$.  Using the unitary freedom (\ref{transformation_meff}), we
bring them into the form
\begin{eqnarray}
\begin{array}{rcl}
m_{\mathrm{eff}}&=& {\rm real,~positive~and~diagonal},\\ [2mm]
Y^{e}&=& {\rm Hermitian}\;.
\end{array}
\label{specbasis1}
\end{eqnarray}
In the non--degenerate case, there is no residual freedom in this
basis due to the Majorana character of $m_{\mathrm{eff}}$. The three
physical phases are therefore
\begin{eqnarray}
\phi_1^{\mathrm{eff}}&=&\mathrm{arg}[Y^e_{12}]\;,\\
\phi_2^{\mathrm{eff}}&=&\mathrm{arg}[ Y^e_{23}]\;,\\
\phi_3^{\mathrm{eff}}&=&\mathrm{arg}[Y^e_{13}]\;.
\end{eqnarray} 

Alternatively, one can choose a basis in which $Y^e$ is diagonal,
\begin{eqnarray}
\begin{array}{rcl}
Y^e &=& {\rm real,~positive~and~diagonal},\\[1.5mm] 
m_{\mathrm{eff}} &=& {\rm symmetric}\;,
\end{array}
\label{specbasis2}
\end{eqnarray} 
where the second equation is satisfied in any basis.
The residual freedom is 
\begin{equation}
\widetilde U_l=\widetilde U_e =\mathrm{diag}(\exp[i\alpha_1],\exp[i\alpha_2],
\exp[i\alpha_3]), \label{phaseredef2}
\end{equation}
such that the three physical phases are of the form
\begin{equation}
{\rm arg} [ (m_{\mathrm{eff}})_{ii}  (m_{\mathrm{eff}})_{jj}  
(m_{\mathrm{eff}})_{ij}^{*2} ] 
\end{equation}
for $i \not=j$.

It is conventional to separate these phases into so-called Majorana
and Dirac ones. This can be done by expressing $m_{\mathrm{eff}}$ as
\begin{equation}
m_{\mathrm{eff}}= U ~ ({\rm real~ diagonal}) ~U^T \;,
\end{equation}
where $U$ is unitary. Five of its phases can be factored out
\cite{dita}
\begin{eqnarray}
U &=& \mathrm{diag}(\exp[i\alpha_1],\exp[i\alpha_2],\exp[i\alpha_3]) \nonumber \\
 &\times & U'~
\mathrm{diag}(1,\exp[i\Phi_1],\exp[i\Phi_2]) \;,
\end{eqnarray}
with $U'$ containing a single phase which cannot be factored out in
this form. The phases $\alpha_{1-3}$ are unphysical and can be
removed by the residual symmetry transformations $ m_{\mathrm{eff}}
\rightarrow\tilde U_l^\dagger m_{\mathrm{eff}}\tilde U_l^*$. The 
``Majorana'' phases $\Phi_{1,2}$ as well as the ``Dirac'' phase
$\delta$ in $U'$ are unaffected by this phase redefinition and are
physical. They enter the PMNS matrix and thus contribute to the
$W$-boson--lepton--lepton vertex \cite{Pontecorvo:1957qd}-\cite{
Pontecorvo:1967fh}.

The necessary and sufficient conditions for CP conservation in the
non--degenerate case are given by
\begin{equation}
\phi_i^{\mathrm{eff}}=0 
\end{equation}
for $i=1,2,3$ which is equivalent to $\Phi_1=\Phi_2=\delta=0$ (the
phases are understood mod $\pi $).

\subsubsection{CP Violating Invariants}

As in the previous subsection, we first construct Hermitian matrices
transforming under one of the unitary symmetries only. At low
energies, $U_l$ is the relevant symmetry and we choose
\begin{eqnarray}
&& {\cal A}= Y^e Y^{e \dagger} \;, \nonumber\\
&& {\cal B}= m_{\mathrm{eff}} m_{\mathrm{eff}}^* \;,\nonumber \\
&& {\cal C}= m_{\mathrm{eff}} (Y^e Y^{e \dagger})^*  m_{\mathrm{eff}}^* \;.
\end{eqnarray}
They all transform as
\begin{equation}
 {\cal M}_i  \rightarrow U_l^\dagger ~ {\cal M}_i ~ U_l \;, 
\end{equation}
where ${\cal M}_i= \{ {\cal A,B,C} \}$.  We first note that generally
${\cal A,B,C} $ are not diagonalizable in the same basis.  Second,
they contain $3\times3-5=4$ invariant phases, three of which are
independent and related to $\phi_i^{\mathrm{eff}}$. Again, using two
Hermitian matrices, \emph{e.g.} ${\cal A}$ and ${\cal B}$, would only
allow us to extract information about a single phase, so it is
necessary to consider ${\cal C}$ as well.

The CP--odd invariants can be chosen as 
\begin{eqnarray}
&&\mathrm{Tr}[ {\cal A,B}]^3 \;,\label{j0'}\\
&&\mathrm{Tr}[ {\cal A,C} ]^3 \;,\label{j1'}\\
&&\mathrm{Tr}([{\cal A,B}] {\cal C})  \;.\label{j2'}
\end{eqnarray}
In the non--degenerate case, they are all independent and can be used
to extract $\phi_i^{\mathrm{eff}}$. This is established by calculating
the Jacobian: $ {\rm Det} \Bigl( \frac{\partial {\cal J}_i }{\partial
\phi_j^{\mathrm{eff}} } \Bigr) $. We thus have three necessary and
sufficient conditions for CP conservation or violation.

As expected, the Jarlskog--type invariant (\ref{j0'}) is independent
of the Majorana phases and is proportional to the Dirac phase,
\begin{equation}
\mathrm{Tr}[ {\cal A,B}]^3 \propto \sin \delta \;.
\end{equation}
It vanishes in the limit of degenerate eigenvalues or
vanishing mixing angles.
The other invariants are
complicated functions of the Dirac and Majorana phases.

The necessary and sufficient conditions for CP conservation in the
non--degenerate case are
\begin{equation}
{\cal J}_i=0 \;,
\end{equation}
where ${\cal J}_i$ ($i=1,2,3$) denote the invariants (\ref{j0'})-(\ref{j2'}).

\subsection{Degenerate Case}

So far we have assumed that there are no degenerate eigenvalues in any
of the matrices and that the mixing angles are non--zero. It is
however instructive to consider the special case, where all the
low-energy neutrino mass eigenvalues are equal, \textit{i.e.} there
exists a basis such that
\begin{equation}
m_{\rm eff}= m \times {\bf 1} \;,
\label{degenerate}
\end{equation}
where ${\bf 1}$ is a $3\times 3$ unit matrix and $m$ is real. In that
case, the special basis (\ref{specbasis1}) is defined up to a real
orthogonal transformation
\begin{equation}
\tilde U_l= \tilde U_e=O ~,~ OO^T={\bf 1} \;, \label{orth}
\end{equation}
which retains the Hermiticity of $Y^e$. Due to this
residual symmetry, the $\phi_i^{\mathrm{eff} }$ are not
all independent and can be parametrized by a single phase
\cite{Branco:1998bw}.

This becomes more transparent in the other special basis
(\ref{specbasis2}), where $Y_e$ is real and diagonal. This basis must
be unitarily related to the basis (\ref{degenerate}) and thus $m_{\rm
eff}$ is given by
\begin{equation}
m_{\rm eff} = m~ U^\dagger_l U_l^* ={\rm symmetric~unitary} \;.
\label{m-eff}
\end{equation}
A symmetric unitary matrix can be parametrized by four phases (and two
angles) \cite{Jarlskog:2005zw}. Indeed, three of them can be factored
out as \cite{dita}
\begin{eqnarray}
&& \mathrm{diag}(\exp[i\alpha_1],\exp[i\alpha_2],\exp[i\alpha_3]) ~~U' ~\times 
\nonumber\\
&& \mathrm{diag}(\exp[i\alpha_1],\exp[i\alpha_2],\exp[i\alpha_3]) \;,
\end{eqnarray}
while the symmetric unitary matrix $U'$ contains a single phase. The
explicit form of $U'$ can be found in \cite{Branco:1998bw}.  The
phases $\alpha_{1-3}$ are removed by the residual phase symmetry
(\ref{phaseredef2}) in this basis, leaving a single physical phase.

Thus, in this degenerate case there is one physical Majorana phase.
This phase has to be Majorana since the Jarlskog invariant Tr$[{\cal
A,B}]^3$ vanishes. [Both $\mathcal{A}$ and $\mathcal{B}$
are diagonal in the basis (\ref{m-eff}).] We observe that the only
non--vanishing invariant is (\ref{j1'}). In the basis where $m_{\rm
eff}$ is diagonal, it is given by (up to a factor)
\cite{Branco:1998bw}
\begin{equation}
{\rm Tr} [ Y^e Y^{e \dagger},  (Y^e Y^{e \dagger})^*]^3 
\end{equation}
and is invariant under the residual orthogonal symmetry (\ref{orth}).
It is non--zero in general since ${\cal A}$ and $ {\cal A}^*$ are
not diagonal in the same basis.

This analysis can be carried over to the ``high energy theory'' case
in a straightforward albeit tedious way.

\section{MSSM with Three Right--Handed 
 Neutrinos \label{MSSM}}

The leptonic part of the most general proton-hexality \cite{
Dreiner:2005rd} (or R-parity) conserving renormalizable superpotential
is given by
\begin{eqnarray}
\mathcal{W}_{\mathrm{leptonic}}&=&-\hat{\mathcal{H}}_2Y^\nu_{ij}\hat L_i
\hat N_j+\hat{\mathcal{H}}_1 Y^e_{ij}\hat L_i
\hat E_j \label{superpot}\\
&&+\frac{1}{2}M_{ij}\hat N_i\hat N_j~.\nonumber
\end{eqnarray}
Here $\hat L$, $\hat E$ and $\hat N$ are the left-chiral superfields
describing the lepton doublet, a charge conjugate of the right--handed
electron and a charge conjugate of the right--handed neutrino,
respectively. $\hat{\mathcal{H}}_1$ and $\hat{\mathcal{H}}_2$ are the
Higgs doublet superfields. The relevant soft SUSY breaking terms are
\begin{eqnarray}
 \Delta V_{\mathrm{soft}}&=&(-\mathcal{H}_2A^\nu_{ij}\tilde l_i \tilde 
n_j^*+\mathcal{H}_1A^e_{ij}\tilde l_i\tilde e_j^*\label{soft-terms}\\
 &&+\frac{1}{2}  B_{ij}\tilde n_i\tilde n_j+\mathrm{H.c.})\nonumber\\
 &&+ M_{ij}^{l\;2} \;\tilde l_i \tilde  l_j^* + M_{ij}^{ \nu\;2}\;\tilde n_i\tilde n_j^*
 +M_{ij}^{ e\;2}\;\tilde e_i\tilde e_j^* \;, \nonumber 
 \end{eqnarray}
 where $\tilde l$, $\tilde e^*$ and $\tilde n^*$ 
are the scalar components of $\hat L$, $\hat E$ and $\hat N$, respectively.
$\mathcal{H}_1$ and $\mathcal{H}_2$ denote the  Higgs doublets. 

As in the SM, the flavour symmetry is 
\begin{equation}
U(3)_l\times U(3)_e\times U(3)_\nu \label{unitarysymmetry} \;,
\end{equation}
which now applies to superfields.\footnote{Fermions and sfermions are
transformed in the same fashion in order to avoid flavour mixing at
the super--gauge vertices.}  The transformation law of the flavour
structures is
\begin{eqnarray}
Y^\nu&\rightarrow&U_l^\dagger ~Y^\nu~ U_\nu~,\label{mssmtrf1}\\
Y^e&\rightarrow&U_l^\dagger ~Y^e~ U_e~,\\
A^\nu&\rightarrow&U_l^\dagger ~A^\nu~ U_\nu~,\\
A^e&\rightarrow &U_l^\dagger ~A^e~ U_e~,\\
M^{ l\;2}&\rightarrow& U_l^\dagger ~M^{ l\;2}~ U_l~,\\
M^{ \nu\;2}&\rightarrow& U_\nu^\dagger ~M^{ \nu\; 2}~ U_\nu~,\\
M^{ e\;2}&\rightarrow& U_e^\dagger ~M^{ e\;2}~ U_e ~,\label{mssmtrf7} \\
M &\rightarrow& U_\nu^T ~M~ U_\nu~,\\
B &\rightarrow& U_\nu^T  ~B~ U_\nu~.
\end{eqnarray}
These objects altogether contain $4\times 9+ 3\times 3 +2\times 6 =
57$ complex phases. The symmetry transformations eliminate $3\times 6$
of them such that we end up with 39 physical CP phases.\footnote{If
the Majorana matrices were absent, we would get $45-17=28$ physical CP
phases.}

In what follows, we classify the corresponding CP phases and CP--odd invariants.

\subsection{SUSY CP Phases and CP--odd Invariants}

In the supersymmetric basis corresponding to (\ref{specbasis}) where
$Y^\nu $ is real and diagonal, and $Y^e$ is Hermitian, the additional
invariant CP phases due to the SUSY flavour structures are given by
\begin{eqnarray}
&& {\rm arg} \Bigl(Y_{ij}^e \; A_{ij}^{ \{e,\nu \} *}\Bigr)\;~~~~~\rightarrow ~18 \;, 
\nonumber\\
&& {\rm arg} \Bigl(Y_{ij}^e \; M_{ij}^{\{e,\nu,l \}\; 2\;*} \Bigr) ~\rightarrow ~9 
\;, \label{susycp}
\\
&& { \rm arg} \Bigl( M_{ij}\; B_{ij}^* \Bigr) ~~~~~~~~~~\rightarrow ~ 6 \;. \nonumber
\end{eqnarray}
These are invariant under the transformations (\ref{phaseredef}).

In the Standard Model, as a next step, we constructed simple Hermitian
objects which \textit{all} transformed under only one of the
symmetries (\ref{tot-sym}). In the MSSM, this approach leads to very
cumbersome expressions. We thus construct three separate groups of
Hermitian objects, which each transform under only one unitary
symmetry, respectively. These are presented in Table~\ref{table1}.  We
find that this set is sufficient to determine all physical phases of
the system in the non--degenerate case.  Before we write down the
CP--odd invariants, let us study what CP phases these Hermitian
matrices are sensitive to.

\begin{table}
\begin{tabular}{|c|c|c|}
\hline
$U(3)_l$ & $U(3)_e$ & $U(3)_\nu$\\
\hline
\hline
$Y^{e} Y^{e\dagger}$ & $Y^{e\dagger}Y^e$ & $Y^{\nu\dagger}Y^\nu$\\
\hline
$Y^{\nu} Y^{\nu\dagger}$  & $A^{e\dagger}A^e$ & $A^{\nu\dagger}A^\nu$\\
\hline
$A^{e} A^{e\dagger}$  & $Y^{e\dagger}A^e+\mathrm{H.c.}$ & $A^{\nu\dagger}Y^\nu+\mathrm{H.c.}$\\
\hline
$A^{\nu}A^{\nu\dagger}$   & $M^{ e\;2}$ & $M^{ \nu \;2}$\\
\hline 
$Y^{e} A^{e\dagger}+\mathrm{H.c.}$  & &$M^*M$\\
\hline
$A^{\nu} Y^{\nu\dagger}+\mathrm{H.c.}$  & &$M^* (Y^{\nu\dagger} Y^\nu)^* M$\\
\hline
$M^{ l\; 2}$ & &$ B^*  (Y^{\nu\dagger} Y^\nu)^*   B$\\
\hline
&&$  B^* M + \mathrm{H.c.}$\\
\hline
\end{tabular}
\caption{The minimal set of  Hermitian flavour objects.\label{table1}}
\end{table}

Consider for example Column~3. In the basis where $Y^{\nu\dagger}Y^\nu
$ is diagonal, the CP phases invariant under the residual symmetry
(\ref{phaseredef}) are of the type
\begin{eqnarray}
&& {\rm arg} ( (\mathcal{M}_i)_{12} (\mathcal{M}_i)_{23}  (\mathcal{M}_i)_{13}^*) 
\;,  \label{phase1b}  \\
&& {\rm arg} ( (\mathcal{M}_i)_{12} (\mathcal{M}_{i+1} )_{12}^*  )\;,..,     
\label{phase2b}
\end{eqnarray}
where $\mathcal{M}_i$ are the Hermitian matrices of the third Column
of Table~\ref{table1}.  Given $N>1$ independent Hermitian matrices,
one can construct $3N-5$ independent invariant phases. These can be
chosen as one CKM--type phase (\ref{phase1b}) and the rest of the form
(\ref{phase2b}).  In this fashion, we obtain 19 invariant phases from
Column~3. However, as we have seen in the SM case, one has to be
cautious in determining the correct number of \textit{independent}
phases, and not too many, since there are certain relations among
these matrices.

In order to make the choice of Hermitian objects in Table~\ref{table1}
plausible and to better understand the counting of independent phases,
consider first the hypothetical special case, when the only non--zero
quantities are $Y^e$, $Y^\nu$ and $M^{\nu\;2}$. In the basis
(\ref{specbasis}) with $M=0$, using the above counting arguments, we
then obtain only four physical independent phases. These \textit{can
not} be recovered from the Hermitian quantities in the three columns
of Table~\ref{table1}. It is only possible to get one phase of the
form (\ref{phase1b}) in Column~1, and another phase of the same type
from Column~3. In order to construct the four phases, it is thus
necessary to include a more complicated Hermitian object,
$Y^{\nu\dagger} Y^e Y^{e\dagger} Y^\nu$, in Column~3, as we did in
Sect.~I. This brings in three extra phases, two of which are
independent. This shows that, in the special case, extra Hermitian
objects may have to be included.

Next let us consider the more involved case, where apart from $Y^e$,
$Y^\nu$ and $M^{ \nu \;2}$, also $A^\nu\not=0$. Again, by our counting
argument, we then have 13 physical independent phases from the
remaining Hermitian objects in Table~\ref{table1} in the
supersymmetric basis corresponding to (\ref{specbasis}). In order to
construct the extra phases, we can now write down additional Hermitian
matrices $A^{\nu}A^{\nu\dagger} $ and $A^{\nu}Y^ {\nu\dagger}
+\mathrm{H.c.}$ in the first column, as well as $A^{\nu\dagger} A^\nu$
and $A^{\nu\dagger}Y^\nu+\mathrm{H.c.}$ in the third column. These
extra objects restore the deficit encountered above, \textit{i.e.} we
can now recover 13 physical phases from the Hermitian objects. The
na\"ive counting gives seven phases for Column~1 and seven phases for
Column~3, which is too many. However, of the matrices
\begin{eqnarray}
&&A^{\nu}A^{\nu\dagger} \;,\; A^{\nu} Y^{\nu\dagger}+\mathrm{H.c.}\;,\;
A^{\nu\dagger}A^\nu \;,\; A^{\nu\dagger}Y^\nu+\mathrm{H.c.} \nonumber
\end{eqnarray}
only three are independent. One of these matrices, say $A^{\nu\dagger}
Y^\nu+\mathrm{H.c.}$, can be reconstructed from the others \cite{
Lebedev:2002wq}. In other words, the nine phases of $A^\nu$ can be
derived from the nine phases of the three Hermitian matrices. This
means that the CKM--type phase associated with $A^{
\nu\dagger}Y^\nu+\mathrm {H.c.}$, namely
\begin{equation}
 {\rm arg}\Bigl(\! ( A^{\nu\dagger}Y^\nu+\mathrm{H.c.})_{12}
 (A^{\nu\dagger}Y^\nu+\mathrm{H.c.})_{23}(A^{\nu\dagger}Y^\nu+\mathrm
 {H.c.})_{13}^* \Bigr)
\end{equation}
is not an independent phase and should not be counted. Although it may
seem that $A^{\nu\dagger}Y^\nu+\mathrm{H.c.}$ should be excluded
altogether, this is not correct since it allows us to restore the
(otherwise missing) phases of $M^{ \nu \;2}$ through the rephasing
invariant combinations
\begin{equation}
{\rm arg} ~\Bigl((M^{ \nu \;2})_{12} (A^{\nu\dagger}Y^\nu+\mathrm{H.c.})_{12}^*  
\Bigr)\;,\;{\it etc.}
\end{equation}
The other three phases can be chosen as
\begin{equation}
{\rm arg} ~\Bigl(( A^{\nu\dagger}A^\nu )_{12} (A^{\nu\dagger}Y^\nu+\mathrm{H.c.})_{12}^*
 \Bigr)\;,\;{\rm etc.}
\end{equation}
We thus end up with six phases from the Hermitian matrices of Column~3
and seven phases from those of Column~1. Similar considerations apply
when adding $A^e$ to Column~2, where the CKM-type phase for
$A^{e\dagger}Y^e+\mathrm{H.c.}$ is not independent.

In the Dirac case, where only $M=B=0$ in (\ref{superpot}),
(\ref{soft-terms}), \textit{i.e.} also $M^l, M^\nu, M^e\not=0$, these
are the only complications and we get 28 phases from the Hermitian
objects of Table~\ref{table1}. Adding a non--trivial Majorana mass $M$
results in five further physical phases.  This is because, in the
basis (\ref{specbasis}), $M$ adds six phases while its overall phase
can be eliminated by the residual symmetry transformation, which
leaves $Y^e$ and $Y^\nu$ invariant. To recover these five phases from
the Hermitian objects, we must add two entries in Column~3, $M^* M$
and $ M^* (Y^{\nu\dagger} Y^\nu)^* M$.  This adds six invariant phases
of the type (\ref{phase2b}), five of which are independent. Finally,
inclusion of $B$ brings in six more physical phases of the type
(\ref{phase2b}) in the basis (\ref{specbasis}), all of which are
independent.  Correspondingly, we add $B^* (Y^{\nu\dagger} Y^\nu)^* B$
and $ B^* M +\mathrm{H.c.}$ to Column~3, which are sensitive to these
phases. Note that the object of the form $ B^* M +\mathrm{H.c.}$ is
necessary as it depends on the $physical$ relative phase between $B$
and $M$. In the end, the first, second and third Column provide 16, 6
and 17 independent phases, respectively.

The above choice of the Hermitian objects is not unique and there are
many other possibilities. In particular, one may replace $A^{\nu
\dagger}A^\nu$ in the third Column with $Y^{\nu\dagger} Y^eY^{e\dagger
} Y^\nu $. In that case, the limit ``soft terms'' $\rightarrow 0$
reproduces the SM Hermitian matrices of
Eqs.(\ref{eq:aequiv}-\ref{eq:aequiv1}). On the other hand, our choice
is similar to the quark sector Hermitian objects of
Ref.\cite{Lebedev:2002wq}. These choices are equivalent in the
non--degenerate case.

The CP--odd invariants are constructed out of the Hermitian objects
transforming under one of the unitary symmetries in
Eq. (\ref{unitarysymmetry}), respectively. These can be chosen
as one Jarlskog--type invariant and the rest $K$--invariants. The
former is sensitive to the cyclic product of phases of a each matrix
while the latter are sensitive to the relative phases between
Hermitian matrices \cite{Lebedev:2002wq}. Thus we have 39 independent
invariants in the non--degenerate case,
\begin{eqnarray}
\begin{array}{l}
J (H_1, H_2) \;, \\
K (H_i^p, H_j^q, H_k^r)\;,
\end{array} \label{jk}
\end{eqnarray}
where $J (A, B) \equiv {\rm Tr} [A,B]^3$,
$K (A,B,C) \equiv {\rm Tr} [A,B,C] $ and $p,q,r$ are integers.
In each invariant, only matrices $H_a$   belonging to  the same column  appear. 
In the Appendix, we give an explicit example of 39 independent invariants.
To prove that they are independent functions of the 39 physical phases 
(\ref{susycp}) and (\ref{phase0}-\ref{phase5}), we have calculated
the Jacobian 
\begin{equation}
{\rm Det} \Bigl(   \frac{\partial J_i}{\partial \phi_j }   \Bigr) \;,
\label{Jsusy}
\end{equation} 
where $J_i$ denotes collectively all the invariants (\ref{jk}) and
$\phi_i$ are the physical phases.  We find that the Jacobian is
non--zero. Thus, all the physical phases can be determined from these
invariants.

We note that the traditional Jarlskog invariants $\mathrm{Tr}[H_i^p,
H_j^q]^r$ {\it are not} sufficient to describe CP violation in
supersymmetry.  This is seen most easily in the case of three
Hermitian matrices $A,B,C$ (which can be, for example, $Y^{e}
Y^{e\dagger}$, $Y^{\nu} Y^{\nu\dagger}$ and $M^{ l\; 2}$).  This
system has four physical phases, however there are only three
independent Jarlskog--type invariants Tr $[A,B]^3$, Tr $[B,C]^3$ and
Tr $[C,A]^3$. All higher order Jarlskog--type invariants are
proportional to these three.  This means that one CP phase cannot be
picked up by such invariants and even if all of them vanish, CP
violation is possible. It is thus necessary to include the $K$--type
invariants
\cite{Lebedev:2002wq}.

The necessary and sufficient conditions for CP--conservation in the
non--degenerate case amount to vanishing of the invariants (\ref{jk}).
In that case, the 39 physical phases vanish and in some basis all the
flavour objects are real. Clearly, there can then be no CP violation
and any higher order CP--odd invariant, e.g. Tr $[A,B,C,D,E,..]$,
would vanish as well.

We will not discuss here the degenerate case in detail. Suffice it to
say that additional conditions such as Im(Tr $(A^e Y^{e\dagger})^n) =
0$, \emph{etc.} arise \cite{Lebedev:2002wq}.\footnote{We are working
under the assumption that different matrices are not diagonal in the
same basis. In the degenerate case, this is not true and all $J$-- and
$K$--invariants can vanish even though there is physical CP
violation. CP--odd invariants sensitive to the corresponding CP phases
are, for example, Tr $[(A^e Y^{e\dagger})^n - {\rm h.c.}]$.}

\subsection{ Low Energy Theory }

Below the seesaw scale $M$, one can integrate out the right--handed
neutrinos as superfields. The resulting theory is the MSSM
supplemented with the dimension-5 operator $\hat{L}\hat{\mathcal{H}}
_2\hat{L}\hat{\mathcal{H}}_2$ (which is proton hexality and R-parity
invariant) generating the left--handed neutrino masses.  The flavour
objects in the low--energy theory are $Y^e$, $m_{\mathrm{eff}}$ and
the soft terms $A^e$, $M^{ l\; 2}$, $M^{ e\;2}$.

In the basis (\ref{specbasis1}), there is no residual rephasing freedom
and the extra SUSY CP phases are
\begin{eqnarray}
&& {\rm arg}(A^e_{ij}) ~~ \rightarrow 9  \;,  \nonumber\\
&& {\rm arg}(M^{ l\; 2}_{ij}) \rightarrow 3  \;,  \\
&& {\rm arg}(M^{ e\; 2}_{ij}) \rightarrow 3  \;, \nonumber
\end{eqnarray}
such that altogether we have 18 physical phases.  The corresponding
basis invariants are built out of the Hermitian matrices of Table
\ref{table2}.  18 independent invariants can be chosen to be of the
form (\ref{jk}) with $H_i$ being the matrices belonging to the same
column of Table \ref{table2}, respectively. Their independence is
established by calculating the Jacobian with respect to the physical
CP phases.  An example of such invariants is given in the Appendix.
The necessary and sufficient conditions for CP--conservation in the
non--degenerate case amount to the vanishing of 18 independent
invariants.

\begin{table}
\begin{tabular}{|c|c|}
\hline
$U(3)_l$ & $U(3)_e$ \\
\hline
\hline
$Y^{e} Y^{e\dagger}$ & $Y^{e\dagger}Y^e$ \\
\hline
$A^{e} A^{e\dagger}$  & $A^{e\dagger}A^e$ \\
\hline
 $Y^{e} A^{e\dagger}+\mathrm{H.c.}$          & $Y^{e\dagger}A^e+\mathrm{H.c.}$ \\
\hline
 $M^{ l\; 2}$     & $M^{ e\;2}$ \\
\hline 
 $ m_{\mathrm{eff}}   m_{\mathrm{eff}}^* $     &    \\
\hline
 $   m_{\mathrm{eff}} (Y^e Y^{e \dagger})^*  m_{\mathrm{eff}}^* $      & \\
\hline
\end{tabular}
\caption{The minimal set of  Hermitian flavour objects in the low energy theory.\label{table2}}
\end{table}

\subsubsection{Observables and CP--odd Invariants }

Physical observables are (complicated) functions of the basis
invariants. An example relevant to CP violation in neutrino
oscillations can be found in \cite{Jarlskog:2004be}.  Here, let us
illustrate this connection with a simple example of the
neutralino--induced electron EDM (see \cite{Abel:2005er} for recent
analyses). In generic SUSY models, it is often expressed in terms of
the ``mass insertion'' $(\delta_{LR}^e)_{11}$ \cite{Gabbiani:1996hi},
\begin{equation}
\Delta  {d_e}\propto  
\mathrm{Im}(\delta_{LR}^e)_{11} \;
\label{edm},
\end{equation}
with 
\begin{equation}
(\delta_{LR}^e)_{11}\approx   \frac{ \langle \mathcal{H}_1  \rangle  
A_{11}^e}{\tilde m^2 } \; ,
\end{equation}
where we have neglected the $\mu$--term contribution.  $\tilde m$ is
the average slepton mass and the $A$--terms are calculated in the
basis where the charged lepton masses are diagonal and real.

To understand the connection to CP--odd invariants, let us assume a
simple form for the $A$--terms in this basis,
\begin{equation}
A^e=\left(\begin{array}{ccc}
A_{11}^e & A_{12}^e & 0\\
0 & 0 & 0 \\
0 & 0 & 0 \\
\end{array}\right) \;.
\end{equation}
Calculating the $K$--invariants with Hermitian matrices of Table
\ref{table2}, Column 2, we find
\begin{eqnarray}
\mathrm{Tr}~\left([Y^{e\dagger}Y^e,(Y^{e\dagger}A^e+\mathrm{H.c.})] 
A^{e\dagger}A^e\right) && \nonumber \\
&& \hspace{-3cm}\propto \sin({\rm arg}~ (A_{11}^e Y_{11}^{e *}))\;.
\end{eqnarray}
We thus conclude that it is this invariant that controls the electron EDM. 

A few comments are in order. First, note the appearance of the
reparametrization invariant phase $ {\rm arg}~ (A_{11}^e Y_{11}^{e
*})$.  Second, this phase cannot be ``picked up'' by any
Jarlskog--type invariant. This is because the $A$--matrix is
effectively 2$\times$2 and the CKM--type phases vanish.  Finally, if
$A_{12}^e =0$, $ A^e$ and $Y^e$ are diagonal simultaneously.  In this
(special) case, the $K$--invariants vanish and CP violation comes from
CP--odd invariants based on anti--Hermitian objects like Tr $[(A^e
Y^{e\dagger})^n - h.c.]$.

In general, even if all of the soft terms are real in some basis, that does not
guarantee absence of dangerous SUSY contributions to EDMs. 
The SM flavour structures  $Y^e$ and $m_{\rm eff}$   may contain complex phases
such that the reparametrization invariant phases are non--zero. In other words,
$K$--invariants can be non--zero even if the soft terms are real. This is 
similar  to the quark sector where the CKM phase can result in large EDMs
in the presence of real soft terms \cite{Abel:2001cv}.

\section{Conclusion}

We have constructed a generalization of the Jarlskog invariant to
supersymmetric models with right--handed neutrinos. We find that CP
violation in supersymmetric models is controlled by CP--odd invariants
of the conventional Jarlskog--type (``$J$--invariants'') as well as
those involving antisymmetric products of three Hermitian matrices
(``$K$--invariants''), which cannot be expressed in terms of the
former.

The presence of right--handed neutrinos brings in new features, in
particular, Majorana--type CP phases in supersymmetric as well as soft
terms. The corresponding CP--odd invariants are built out of Hermitian
objects involving a product of two or four flavour matrices as opposed
to 2 in the Dirac case.  This complicates the analysis, on the one
hand, but allows for interesting features, on the other hand. For
example, CP violation is possible even if the neutrinos are all
degenerate in mass.

We have identified 39 physical CP phases and corresponding CP--odd
invariants which control CP violation in the lepton sector of the MSSM
with right--handed neutrinos. Below the seesaw scale, the low energy
theory is described by 18 CP phases which can again be linked to 18
independent CP invariants.  This allows us to formulate
basis--independent conditions for CP conservation in the
non--degenerate case.

Physical observables are in general complicated functions of CP--odd
invariants, which we illustrate with an example of the electron
EDM. SUSY CP violation and, in particular, dangerous EDM
contributions, are possible even if the soft supersymmetry breaking
terms are real in some basis.

\section*{Acknowledgements}
We thank Howie Haber for helpful discussions. M.T. greatly appreciates
that he was funded by a Feodor Lynen fellowship of the Alexander von
Humboldt foundation, and he also thanks the Physikalisches Institut in
Bonn for hospitality.

\begin{appendix}
\section{Independent CP--odd invariants}

Let us label  matrices of the first column of  Table \ref{table1}
by $X_i$, second -- $Y_i$, and third -- $Z_i$,
where $i$ refers to the row number. Then the 39
independent invariants can be chosen as  
\begin{eqnarray}
&&\mathrm{Tr}[X_1,X_2]^3,\\
&&\mathrm{Tr}[X_1,X_2]X_3,\\
&&\mathrm{Tr}[X_1^2,X_2]X_3,\\
&&\mathrm{Tr}[X_1,X_2^2]X_3,\\
&&\mathrm{Tr}[X_1,X_2]X_4,\\
&&\mathrm{Tr}[X_1^2,X_2]X_4,\\
&&\mathrm{Tr}[X_1,X_2^2]X_4,\\
&&\mathrm{Tr}[X_1,X_2]X_5,\\
&&\mathrm{Tr}[X_1^2,X_2]X_5,\\
&&\mathrm{Tr}[X_1,X_2^2]X_5,\\
&&\mathrm{Tr}[X_1,X_2]X_6,\\
&&\mathrm{Tr}[X_1^2,X_2]X_6,\\
&&\mathrm{Tr}[X_1,X_2^2]X_6,\\
&&\mathrm{Tr}[X_1,X_2]X_7,\\
&&\mathrm{Tr}[X_1^2,X_2]X_7,\\
&&\mathrm{Tr}[X_1,X_2^2]X_7.
\end{eqnarray}

\begin{eqnarray}
&&\mathrm{Tr}[Y_1,Y_3]Y_2,\\
&&\mathrm{Tr}[Y_1^2,Y_3]Y_2,\\
&&\mathrm{Tr}[Y_1,Y_3^2]Y_2,\\
&&\mathrm{Tr}[Y_1,Y_3]Y_4,\\
&&\mathrm{Tr}[Y_1^2,Y_3]Y_4,\\
&&\mathrm{Tr}[Y_1,Y_3^2]Y_4.
\end{eqnarray}

\begin{eqnarray}
&&\mathrm{Tr}[Z_1,Z_3]Z_2,\\
&&\mathrm{Tr}[Z_1^2,Z_3]Z_2,\\
&&\mathrm{Tr}[Z_1,Z_3^2]Z_2,\\
&&\mathrm{Tr}[Z_1,Z_3]Z_4,\\
&&\mathrm{Tr}[Z_1^2,Z_3]Z_4,\\
&&\mathrm{Tr}[Z_1,Z_3^2]Z_4,\\
&&\mathrm{Tr}[Z_1,Z_3]Z_5,\\
&&\mathrm{Tr}[Z_1^2,Z_3]Z_5,\\
&&\mathrm{Tr}[Z_1,Z_3^2]Z_5,\\
&&\mathrm{Tr}[Z_1,Z_3]Z_6,\\
&&\mathrm{Tr}[Z_1^2,Z_3]Z_6,\\
&&\mathrm{Tr}[Z_1,Z_3]Z_7,\\
&&\mathrm{Tr}[Z_1^2,Z_3]Z_7,\\
&&\mathrm{Tr}[Z_1,Z_3^2]Z_7,\\
&&\mathrm{Tr}[Z_1,Z_3]Z_8,\\
&&\mathrm{Tr}[Z_1^2,Z_3]Z_8,\\
&&\mathrm{Tr}[Z_1,Z_3^2]Z_8.
\end{eqnarray}

Similarly, labelling  entries of the first column of  Table \ref{table2} 
by $A_i$ and those of the second column by $B_i$,
we have the following 18 independent
invariants:
\begin{eqnarray}
&&\mathrm{Tr}[A_1,A_6]^3,\\
&&\mathrm{Tr}[A_5,A_1]A_6,\\
&&\mathrm{Tr}[A_5^2,A_1]A_6,\\
&&\mathrm{Tr}[A_5,A_1]A_2,\\
&&\mathrm{Tr}[A_5^2,A_1]A_2,\\
&&\mathrm{Tr}[A_5,A_1^2]A_2,\\
&&\mathrm{Tr}[A_5,A_1]A_3,\\
&&\mathrm{Tr}[A_5^2,A_1]A_3,\\
&&\mathrm{Tr}[A_5,A_1^2]A_3,\\
&&\mathrm{Tr}[A_5,A_1]A_4,\\
&&\mathrm{Tr}[A_5^2,A_1]A_4,\\
&&\mathrm{Tr}[A_5,A_1^2]A_4,
\end{eqnarray}

\begin{eqnarray}
&&\mathrm{Tr}[B_1,B_3]B_2,\\
&&\mathrm{Tr}[B_1^2,B_3]B_2,\\
&&\mathrm{Tr}[B_1,B_3^2]B_2,\\
&&\mathrm{Tr}[B_1,B_3]B_4,\\
&&\mathrm{Tr}[B_1^2,B_3]B_4,\\
&&\mathrm{Tr}[B_1,B_3^2]B_4.
\end{eqnarray}

\vfill

\end{appendix}

%
%%%%%%%%%%%%%%%%%%%%%%%%%%%%%%%%%%%%%%%%%%%%%%%%%%%%%%%%%
%

\end{document}